\documentstyle[prd,aps,preprint,12pt]{revtex}

\tightenlines

\begin{document}
\newcommand{\nn}{\nonumber\\}
\newcommand{\p}[1]{(\ref{#1})}

\newcommand{\be}{\begin{equation}}
\newcommand{\ee}{\end{equation}}
\newcommand{\bea}{\begin{eqnarray}}
\newcommand{\beas}{\begin{eqnarray*}}
\newcommand{\eea}{\end{eqnarray}}
\newcommand{\eeas}{\end{eqnarray*}}
\newcommand{\ba}{\begin{array}}
\newcommand{\ea}{\end{array}}
\renewcommand{\thefootnote}{\fnsymbol{footnote}}

\newcommand{\e}{\epsilon}
\newcommand{\ka}{\kappa}
\newcommand{\n}{\eta}



\preprint{\vbox{\hbox{ OSU-HEP-01-12}}}

\title{Small Dirac Neutrino Masses and $R$-parity from
Anomalous $U(1)$ Symmetry}

\author{Ilia Gogoladze $^{a,b}$\footnote{On a leave of absence from:
Andronikashvili Institute of Physics, GAS, 380077 Tbilisi, Georgia.\\
iliag@ictp.trieste.it}
 and
Abdel P\'erez-Lorenzana$^{a,c}$\footnote{aplorenz@ictp.trieste.it}}
\address{
$^a$ The Abdus Salam International Centre for Theoretical Physics,
     I-34100, Trieste, Italy\\
$^b$ Department of Physics, Oklahoma State University,
     Stillwater, OK 74078, USA\\
$^c$ Departamento de F\'{\i}sica,
Centro de Investigaci\'on y de Estudios Avanzados del I.P.N.\\
Apdo. Post. 14-740, 07000, M\'exico, D.F., M\'exico}

\date{November, 2001}

\maketitle

\begin{abstract}
We suggest that  many of the free parameters in the supersymmetric
extensions of the Standard Model can be all linked together to the
existence of a non universal $U(1)$ gauge symmetry, which has been
spontaneously broken at very high scale. Such a symmetry can
easily generate, via non-renormalizable operators, appropriate
tree level fermion mass textures as well as the $\mu$-term of the
Higgs potential. We give a general parametrization of those terms.
As an output,  R parity breaking terms only appear at a
non-renormalizable level and linked to the Yukawa couplings,
giving rise to the possibility of having an effective (exact) R
parity conservation. As an interesting application of this idea we
explore the case where neutrinos are Dirac particles. The scenario
can be embedded in an extended $SU(5)$ unification theory where
the extra right handed neutrinos are introduced as singlets. Such
theory has an exact R parity conservation and sneutrino as LSP.
\end{abstract}

\section{Introduction}

Supersymmetry (SUSY)~\cite{susy} has been considered for many
years as the  preferred solution to the hierarchy problem that
pollutes the  Higgs sector in the Standard Model (SM) of particle
physics. The natural cancellation of quadratic divergences in the
SUSY theory, which takes place among particles and sparticles
belonging to same representation, makes it a very appealing
extension of the SM. However, as it is well known, SUSY is not an
exact symmetry in nature but rather a broken one, and  the
effective SUSY  SM  has more than one hundred free
parameters. About twenty  of these parameters correspond to those
originally contained in the SM, among which one has masses and
mixings of fermions, three coupling constants, a CP phase and the
$\mu$ coupling on the Higgs potential. Many other of those
parameters are the so called soft breaking terms, which encode our
ignorance of the way SUSY is being broken at the high scale. In
the SUSY limit, however, new renormalizable gauge invariant terms,
which explicitly violate lepton and baryon number conservation,
can also be written down. Usually such terms are dangerous since
they may give rise to a fast proton decay, or to flavour changing
neutral processes at large rates. The usually advocated solution
of this problem  is the addition of an extra discrete symmetry,
called R parity,  which distinguishes particles from their
superpartners, and which also  forbids such dangerous operators.
On the contrary, in the absence of such a symmetry one needs to
introduce small couplings which can be sources of new phenomena.
An interesting use of R-parity breaking terms could be, for
instance, the generation of neutrino masses~\cite{rpnu}. Although
in the SM, the neutrino is massless, recent experimental data
points towards massive neutrinos (with masses smaller than few eV)
and to one or possibly two large mixing angles in the lepton
sector~\cite{sol,atm}. The situation is then opposite to that in
quark sector where mixing angles are small.

With  massive neutrinos there also comes the  challenging problem
of understanding whether neutrinos are Majorana or Dirac
particles. So far, no evidence that may resolve this question
exists. A positive signal in future  neutrinoless double beta
decay experiments would decide for a Majorana neutrino. However, a
negative result would not shed any light up on this problem. From
the theoretical point of view, the generation of neutrino masses
usually needs extending the SM particle content by adding  right
handed neutrinos $N_R^i$ (where $i=1,2,3$ is the family index),
which can be either SM singlets or triplets under $SU(2)_L$ gauge
symmetry.  Notice, however, that  (Majorana) neutrino masses can
still be generated even without introducing extra fields (see for
instance Ref.~\cite{rpnu}). With a right handed singlet we can
write the following Yukawa couplings:
\begin{equation}
\label{dir}
 h_{ij}L_{i} H_{2} N_R^j~,
\end{equation}
where $L_i$ is left handed doublet, $H_2$ is SM Higgs field and
$h$ are the Yukawa couplings. Once Higgs field gets a VEV,
Eq.~(\ref{dir}) induces Dirac masses for the neutrinos,
$M_D L_{i} N_R^i$. On the other hand, a priori one can not forbid
Majorana mass terms for right handed neutrinos, $M_R \bar
N_R^cN_R$. Thus, the question of whether the neutrino is a
Majorana or a Dirac field clearly depends on the $M_R$ scale. So,
we  define a neutrino to be a:
\begin{enumerate}
\item Majorana particle if $M_R \gg M_D$;
\item Pseudo-Dirac particle if  $M_R \ll M_D$;
\item Dirac particle if $M_R=0$.
\end{enumerate}
Last two cases are difficult to accommodate theoretically since
they may required  a large fine tuning on the Yukawa couplings to
get eV masses. In fact one needs $h\sim O(10^{-11})$ or less. Some
models that attempt to explain the smallness of this coupling can
be found in Refs.~\cite{d1,d2,d3}. In contrast, the lightness of
neutrino as a Majorana particle is easier to explain through the
seesaw mechanism~\cite{seesaw}. For this reason neutrinos as
Majorana particles are usually considered as the most appealing
possibility.

In any case, the extended theory that includes an explanation for
neutrino anomalies will contain at least six more parameters:
three neutrino masses and their three mixing angles. The sole
existence in the theory of all those (fundamental) parameters has
made us to believe that neither the SM, nor its minimal SUSY
version (MSSM), can be the final theory of particle physics.
Understanding their origin has motivated the exploration of many
new ideas and models for particle physics at higher energies.
Examples of it are the extended gauge models as the Left-Right
models and the Grand Unified Theories (GUT). It is expected that
some of these ideas may find a realization in String Theories. The
usual trend in the study of this problem, however, is to analyze
each sector, or set of parameters, in an almost independent
way. The pattern of neutrino masses  is usually studied separately
from all other sectors and in fact new scales, and so new physics,
are usually introduced~\cite{seesaw}. GUT explains the values of
the coupling constants but it can say very little about the masses
and mixings among fermion families without the help of extra
(flavour) symmetries, though some links between lepton and quark
mixing and masses can still be obtained~\cite{paty}. Indeed, using
Abelian flavor symmetries~\cite{ab,joshi} to understand the mass
hierarchies is very popular.

In this paper we are re-addressing the question of whether all
these ingredients of the theory may all be linked together through
the existence of some Abelian $U(1)_A$ symmetry. The overall idea
is as follows. We assume that there exists an extra symmetry that
can distinguish among the different MSSM particle representations.
Thus, except for gauge and soft SUSY breaking couplings, all other
couplings (Yukawa; $\mu$ and R-parity breaking terms) might not be
allowed at the renormalizable level by $U(1)_A$ gauge invariance.
In fact, $U(1)_A$ charges might be such that only certain
couplings, among all, would be acceptable. As there are not too
many degrees of freedom (total number is equal to number of field
representations), once we fix them by asking, for instance, for
specific profiles (textures) of the lepton and quark mass
matrices, all other allowed couplings of the theory would be
predicted. Moreover, if our $U(1)_A$ is broken at some large
energy scale by a SM singlet, then, those operators that were not
present in the unbroken theory can now be generated by the use of
non-renormalizable operators. Such operators come with a
suppression respect to the renormalizable ones, thus they can be
responsible for generating the observed hierarchy on the masses,
as well as  suppressing (forbidding) R-parity breaking terms.
However, interestingly enough, this mechanism can only generate a
very specific class of mass matrices for the minimal matter
content. Although in our mechanism one can equally use global as
well as local symmetries, we will rather prefer to use local ones.
An interesting example of such a theory is the case of anomalous
gauge symmetries~\cite{ab,joshi}. Such anomalous $U(1)_A$
symmetries may be related to String Theory. They are usually
broken close to string scale and the anomaly canceled by
Green-Schwarz~\cite{gs} mechanism. 
Note that the freedom that an Abelian 
symmetry has on charge assignments may be reduced if one
considers non-Abelian symmetries instead. 
Alternative models of these sore can be found on Ref.~\cite{rasin}.

We organize the present discussion as follows. First, to motivate
the ideas, we briefly review the Green-Schwarz~\cite{gs} mechanism
used to cancel the $U(1)_A$ anomaly. Then, we shall discuss the
general procedure used to determine the mass matrices induced by
the $U(1)_A$ symmetry. As we will show, the mass textures
generated by this mechanism can be  parameterized by a small
number of parameters, which encode the  actual charge assignments
on matter fields. As a direct application of our whole idea, we
study the possibility of using an Abelian $U(1)_A$ symmetry for
getting an effective theory that accommodates Dirac neutrinos. We
also show that R-parity breaking terms are linked to the effective
Yukawa couplings. In fact in a theory with Dirac neutrinos only
baryon number violating interactions would be present. We further
explore the possibility that such a theory can be embedded in a
GUT environment, particularly in an extended $SU(5)$ theory.
There, one gets a theory that predicts exact R-parity
conservation. Thus, LSP appears as a dark matter candidate, which
in our example could correspond to the right handed sneutrino.
Some explicit examples of tree level  mass matrices  are given at
the end of the paper.


\section{Framework}

Let us start by discussing the general mechanism  we want to use
for protecting mass textures and constraining R-parity breaking
terms. In this paper we will always work in the context of SUSY.
In order to allow for (Dirac) neutrino masses we extend the MSSM
by introducing  three right handed neutrino superfields, $N_i$. In
addition, we introduce a $U(1)_A$ flavor symmetry, which
distinguishes the different matter representations of the MSSM
through the prescription of $U(1)_A$ charges. We will also assume
that this symmetry has been  broken spontaneously, at some high
scale, by a heavy scalar superfield S. Given this, our working
philosophy will be to consider, at tree level, all those
renormalizable and non-renormalizable operators that are allowed
for the theory. Among them, we will pay most of our attention to
those which, up on symmetry breaking, reproduce Yukawa couplings,
$\mu$ and  R-parity breaking terms.  Therefore, we will take the
approach that try to understand the hierarchies as a consequence
of large suppressions coming on high dimension operators. Our
final goal will be parameterizing the most general form of all
those couplings in the effective low energy theory.
It is worth mentioning that the power of this effective field theory 
approach is that it gives a well defined framework for knowing where the small
numbers appear on the theory. 
However, we should clearly state that 
we will not provide a dynamical model 
that would yield exactly the set of higher dimensional operators that 
we will discuss along the paper.

In the absence of any other heavy particles, it turns out that the
$ U(1)_A$ symmetry is anomalous. It is known that such anomalous $
U(1)_A$ factors can arise in string theories. Cancellation of such
anomaly occurs through the Green-Schwarz mechanism~\cite{gs}.
Though, we will assume so for the moment, this is an additional
ingredient which may not be at all required for our mechanism,
since heavy exotic matter may also take care of balancing the
theory, for instance. Due to the  Abelian $U(1)_A$
symmetry, a Fayet-Illiopoulos D-term $\xi \int\! d^4\theta~V_A$
is always generated, where in string theory $\xi $ is given
by~\cite{fi}
 \begin{equation}
 \xi =\frac{g_A^2M_P^2}{192\pi^2}~{\rm Tr}~Q~.
 \label{xi}
 \end{equation}
The $D_A$-term will have the form:
 \begin{equation}
 \label{da}
 \frac{g_A^2}{8}D_A^2=\frac{g_A^2}{8}
 \left(\Sigma Q_a|\varphi_a |^2+\xi \right)^2~,
 \end{equation}
where $Q_a$ is the `anomalous' charge of $\varphi_a $ superfields.
In order to break down $U(1)$ we introduce a singlet superfield
$S$ with  $U(1)$ charge $Q_S$.
Assuming $\xi>0~$ [${\rm Tr}Q>0$ in (\ref{xi})], and taking $Q_S=-1~$, the
cancellation of $D_A$ fixes the VEV of the scalar component of $S$
field:
 \begin{equation}
 \label{ss3}
 \langle S\rangle =\sqrt{\xi }.
 \end{equation}
Furthermore, we will take
 \begin{equation}
 \frac{\langle S \rangle }{M_P}\equiv \epsilon \simeq 0.2~.
 \label{epsx}
 \end{equation}
As we shall see bellow,  $\epsilon $ turns out to be an important expansion
parameter. Indeed due to the coupling of $S$ with all other MSSM fields,
$\epsilon $ parameter governs all mass textures  in the
theory, as we will now discuss. From this point of view,
fermion masses and mixings
would be  the residual witnesses of the existence of the $U(1)_A$ symmetry.

\section{Quark Textures from  a $U(1)_A$ symmetry}

Lets consider the following general assignment
of $U(1)_A$ charges to the MSSM and right handed neutrino
field representations,
\be
\ba{lll}
L_i (1,2,-1/2)~~:~\alpha_i~~;\qquad
&N_i(1,1,0)\qquad\, :~\beta_i~~;&\qquad
E^c_i (1,1,1)\quad~:~\delta_i ~~;\\
Q_i(3,2,1/6)~~~~:~\theta_i~~;\qquad
&U^c_i(\bar{3},1,-2/3)~:~\omega_i~~;&\qquad
D^c_i({\bar 3}, 1,1/3)~:~\rho_i~~; \\
H_u(1,2,1/2)\quad:~ \gamma~~;\qquad&
H_d(1,2,-1/2)~:~ \sigma~~, &
\ea
\label{charges}
\ee
where  in the brackets  are  given the
$SU(3)_c\times SU(2)_L\times U(1)_Y$ quantum numbers of the particle.
Also, The index $i$ stands for family replication. With these
choices, Yukawa couplings are not trivially allowed by gauge
invariance. Instead, the superpotential contains a  class of
operators that involve couplings to  $S$ field.

In the up quark sector, for instance, the most general couplings
that are trilinear in the MSSM fields have the form
 \be
 W_u = \sum_{ij} h_u^{ij}~Q_i H_u U^c_j \left({S\over M_P}\right)^{n_{ij}}~,
\label{wu}
 \ee
where $n_{ij}$ is a set of integer numbers.  Accordingly to our
working philosophy, the dimensionless coupling constants, $h_u$,
would be taken hereon to be of order one. 
Notice that this apparent extra degrees of freedom would not actually 
play any important role on our understanding of small numbers on the theory. 
Indeed, such order one parameters are not really sizeable, 
nor arbitrarily adjustable, as to account for the features 
of the mass spectrum, and so we will  omit them along our further discussion, 
though their presence should be understood.
After the scalar part of the
$S$ superfield has developed its VEV, the above superpotential
will give rise to the effective Yukawa couplings
 \be
  Y_u^{ij} = \epsilon^{n_{ij}}~,
\label{yu}
 \ee
which generate the up quark mass matrix $M_u \sim Y_u~v$, with $v$
the Higgs VEV. Therefore,  we can
easily get a hierarchical mass pattern defined in terms of powers
of a single parameter, $\epsilon$, just as desiderated in many
cases. We shall stress, however, that not just any arbitrary
texture can be accommodated through this mechanism. Indeed, gauge
invariance constrains the exponents by relating them to the
charges of the MSSM fields in Eq.~(\ref{charges}), such that
 \be
 n_{ij}=  \theta_i + \omega_j + \gamma~.
\label{nchar}
 \ee
Let us mention for completeness that a negative (or non integer)
$n_{ij}$ would actually indicate a forbidden coupling. It is then
straightforward to show that the above formula translates into a
symmetry relationship among the $n_{ij}$ parameters. Indeed, for
any four given family indexes one gets
 \be
 n_{ij} + n_{kl} = n_{kj} + n_{il}~.
\label{nn}
 \ee
Eq. ~(\ref{nn}) reduces the number of linearly independent
$n_{ij}$ parameters to five in a three family basis. An easy way
to understand this point is as follows. First notice that
Eq.~(\ref{nn}) relates the exponents of the four elements of any
two by two submatrix of $Y_u$. Therefore, only three of those
elements are actually independent. One actually gets \be
\left(\ba{cc}
         \e^{n_{ij}} & \e^{n_{il}} \\
         \e^{n_{kj}} & \e^{n_{kj} + n_{il} - n_{ij}}
     \ea \right)~.
\ee Now we proceed by adding a row (or a column) to the above
matrix. Furthermore,  Eq.~(\ref{nn}) will constrain one of the two
elements of such a row (column),  and so, our degrees of freedom
get increased only by one. By induction we are then led to the
conclusion that in a model with $N$ flavours the Yukawa couplings
of Eq.~(\ref{yu}) are defined by  $2N -1$ degrees of freedom.
Thus, for $N=3$ one has only five.

A similar conclusion will follow from considering the down quark couplings
\be
 W_d = \sum_{ij} Q_i H_d D^c_j \left({S\over M_P}\right)^{k_{ij}}~,
\label{wd}
 \ee
where now $k_{ij}$ represents the power on $\epsilon$
suppression that appears in the
effective Yukawa couplings
$Y_d^{ij} = \epsilon^{k_{ij}}$.
We shall notice that besides the already known constraint among
the $k_{ij}$ parameters,
given by exchanging $n\rightarrow k$ in Eq.~(\ref{nn}),
one also gets a new constraint that interrelates both, up and down, sectors,
due to the fact that both Eqs.~(\ref{wu}) and (\ref{wd}) share
a common field, $Q$. By combining the relation
$k_{ij}=  \theta_i + \rho_j + \sigma$
with that in Eq. (\ref{nchar}) one easily gets
\be
 n_{ij} + k_{kl} = n_{kj} + k_{il}~,
 \label{nk}
\ee
for any arbitrary set of family indexes $i,j,k,l$.

Once we have fixed all five
free parameters that determine the up mass texture, above constraints
will leave only three more independent  parameters to define the down mass
matrix.
Therefore,
the most general Yukawa couplings that one can build from our
mechanism can be written as
\be
 Y_u \sim \left(\ba{ccc}
         \e^{n_3 + n_5} & \e^{n_2 + n_5} &\e^{n_5}\\
         \e^{n_3 + n_4} & \e^{n_2 + n_4} &\e^{n_4}\\
         \e^{n_3 } & \e^{n_2} & 1
       \ea\right) \e^{n_1}~; \qquad\mbox{ and }\qquad
 Y_d \sim \left(\ba{ccc}
         \e^{k_3 + n_5} & \e^{k_2 + n_5} &\e^{n_5}\\
         \e^{k_3 + n_4} & \e^{k_2 + n_4} &\e^{n_4}\\
         \e^{k_3 } & \e^{k_2} & 1
       \ea\right) \e^{k_1}~;
\label{yq}
 \ee
where $n_{1,...,5}$ and $k_{1,2,3}$ represent the eight arbitrary
 integer numbers that parametrize our (model dependent) degrees
of freedom. Notice that we have factorized the overall
suppressions, $\e^{n_1}$ and $\e^{k_1}$, which characterize the
mass scales of quarks in third family.

\section{Lepton Textures.}

We now turn our attention to the lepton sector. We then consider the
following interactions
 \be
 W_{\ell} =\sum_{ij}\left[
   L_i H_u N_j \left({S\over M_P}\right)^{p_{ij}} +
   L_i H_d E^c_j \left({S\over M_P}\right)^{q_{ij}}\right]~,
 \label{wl}
 \ee
which give rise to the effective
Yukawa couplings, $Y_\nu^{ij}=\e^{p_{ij}}$ and
$Y_e^{ij}=\e^{q_{ij}}$, that are
responsible for the generation of masses for
Dirac neutrinos and charged leptons, respectively.
Once again, the analysis
follows  same lines as in the quark sector above.
$p_{ij}$ and $q_{ij}$ are related to  $U(1)_A$ charges
by $p_{ij} =  \alpha_i + \beta_j +\gamma$ and
$q_{ij} = \alpha_i + \delta_j+ \sigma $.
By combining this expressions
we will find, again, that only eight free
parameters govern the profiles in the mass matrices,
$M_{e,\nu} = Y_{e,\nu}~v$, also called  the mass textures.
We then choose those parameters in order
to write the most  general lepton  textures in the useful form
\be
 Y_\nu \sim \left(\ba{ccc}
         1         & \e^{p_2 }    &\e^{p_3}\\
         \e^{p_4}  & \e^{p_2+p_4} &\e^{p_3+ p_4}\\
         \e^{p_5 } & \e^{p_2+p_5} &\e^{p_3+ p_5}
       \ea\right) \e^{p_1}~; \qquad\mbox{ and }\qquad
 Y_e \sim \left(\ba{ccc}
         \e^{q_3-p_5}      & \e^{q_2-p_5}     &\e^{-p_5}\\
         \e^{q_3+p_4-p_5}  & \e^{q_2+p_4-p_5} &\e^{p_4-p_5}\\
         \e^{q_3 }         & \e^{q_2}         & 1
       \ea\right) \e^{q_1}~,
\label{yl}
 \ee
In this expressions we have
used five $p$ integer numbers to parametrize the
Dirac neutrino couplings, and  three more, $q$,
for $Y_e$. Notice that the matrices in Eq.~(\ref{yl}) are
just a reparametrization of
those already given for the quark sector in Eq.~(\ref{yq}).
As before, the  factors $\e^{p_1}$ and $\e^{q_1}$
give overall suppression scales. However, the larger scale in each
sector may be different, since it actually
depends on our specific choice for the  numerical values of the $p$ and $q$
parameters.

There will also be some operators, which are bilinear in the low energy fields,
that induce Majorana mass terms for the right handed neutrinos. They are
written as
\be
  W_{R} =\sum_{ij} M_P~N_iN_j \left({S\over M_P}\right)^{r_{ij}}~,
 \ee
with $h_R$ of the order of one, and
$r_{ij} = \beta_i + \beta_j$.
Therefore, the induced Majorana masses will have the form
$(M_R)_{ij}=Y_R^{ij}~M_P$ where $Y_R^{ij}= \e^{r_{ij}}$.
A priori, and due to the symmetry in $r_{ij}$,
the right handed mass terms should be
defined in terms of three  different $r$ parameters.
However, once we have fixed the
Dirac textures in Eq.~(\ref{yl}), there would be
only one extra degree of freedom that one can
use to do the parametrization. One can easily understand this fact
by noting that there is a constraint that relates the $p_{ij}$ parameters
on the $Y_\nu$ texture with the $r_{ij}$ ones of $Y_R$,
which is similar to that of Eq.~(\ref{nk}).
Indeed, from gauge invariance one implies
$ p_{ij} + r_{lk} = p_{ik} + r_{lj}$.
Thus, we can parametrize the right handed masses as
\be
 M_R \sim \left(\ba{ccc}
         1 & \e^{p_2 } &\e^{p_3}\\
         \e^{p_2} & \e^{2 p_2} &\e^{p_2+ p_3}\\
         \e^{p_3} & \e^{p_2+p_3} & \e^{2 p_3}
       \ea\right) \e^{r}~M_P~.
\label{mr} \ee By looking at $M_R$ we notice that  since we have
chosen $p_{1,2,3}$ to be integer numbers, there is a possibility
of forbidding all Majorana masses just by taking $r$ to be either
a fractional  or a large  negative number. This is an interesting
possibility that we want to explore in the last part of the paper.
On the other hand, a large and positive $r$ would add for a large
suppression on $M_R$, thus leading to pseudo-Dirac neutrinos.

Among all other non-renormalizable operators that  low energy
theory would have, we will find those of the form
\be
\e^{s_{ij}}{L_i H_u L_j H_u \over M_P}~;
\label{ll}
\ee
and
 \be
\e^{t_{ij}}{N_i N_j H_u H_d \over M_P}~.
\label{mn}
\ee
In principle, they may induce tiny Majorana masses for left and
right handed neutrinos. In the case where $M_R$ is large, we can
safely neglect these two last contributions. However, in case
$M_R$ terms were forbidding, it is not obvious that same holds for
these two operators. If they were allowed they would split the
Dirac neutrino  components into a pseudo-Dirac pair. The induced
mass squared difference could be too small as to contribute to
neutrino oscillations. In this case it is easy to show that the
exponents in the class of terms of Eq.~(\ref{ll}) are totally
fixed by the parameters already used for the other lepton
textures. In fact, after some simple algebra one gets
$s_{ij}=p_{i1} + p_{j1} - r$, which  also means that
 \be
 \e^{s_{ij}}\sim Y_\nu^{i1} Y_\nu^{j1} \e^{-r}
 \ee
Clearly, if $r$ is chosen to be a  non integer number (or a large
enough negative one), then such terms would  be  forbidden.  This
is it due to our choice on the $p$ parameters as integer numbers.
We shall postpone the discussion of the remaining terms for the
next section, since they are closely related to the $\mu$-term
that we will now discuss.

\section{$\mu$-term}

It is worth stressing that above textures have been calculated yet
with out  knowing the actual values of $U(1)_A$ charges in
Eq.~(\ref{charges}). In fact, our whole procedure  has been rather
to express such charges in terms of the above given parameters.
Thus, all our low energy physics could be written in terms of a
set of integers numbers with a clear physical meaning, that of the
hierarchies. One should notice, however, that so far we have
involved up to twenty different charges, corresponding to equal
number of superfield representations, which were used in writing
down the superpotential terms $W_u$, $W_d$, $W_\ell$ and $W_R$.
Nevertheless, only seventeen of them can be fixed by any given
choice of Yukawa textures. Thus, at least three of all will remain
undefined at this point.

One extra constraint comes from the $\mu$-term, which is again
non trivially allowed at the renormalizable level 
(unless of course $\gamma =- \sigma$ which we do not assume). 
Indeed, the most general
coupling that induces a $\mu$-term on the effective low energy theory is
given, up to an order one coupling constant, as
 \be
  M_P~H_u H_d\left({S\over M_P}\right)^a~,
\label{mut}
 \ee
for $a$ an integer number that satisfies $a = \gamma + \sigma$.
Hence, by setting in the VEV of the $S$ field, one gets the
induced $\mu H_u H_d$ term, where
 \be
 \mu = \e^a M_P~.
 \ee
The large hierarchy between Planck and electroweak scale
fixes $a =\ln(\mu/M_P)/\ln\e \sim 16$.
We then choose
$a$ as a parameter of the  theory.

Notice that the  coupling (\ref{mut})
also enters in the operator mentioned in Eq.~(\ref{mn}),
which generates tiny right handed neutrino masses. The possible
texture of those terms would be completely fixed by that in $Y_R$. Indeed,
one gets
\be
\e^{t_{ij}}  \sim \left({\mu\over M_P^2}\right)~ (M_R)_{ij}~.
\ee
Therefore, the couplings in Eq.~(\ref{mn})
appear just as a correction to any already
existing right handed mass term.
Hence, forbidding $M_R$  will also forbid these extra terms.
We shall notice, however,
that in the special case where one forbids $M_R$
couplings by taking $r$ to be negative,
the mass terms in Eq.~(\ref{mn}) could still be generated
due  to possible conspiring  cancellations among the parameters, which may
render $a + r_{ij}>0$. In such a case the theory would
generate pseudo-Dirac neutrinos.

\section{R-parity breaking interactions}

Last condition imposed by the $\mu$-term reduces our ignorance
on the charge assignments to only two undefined degrees of freedom.
In the absence of new (model dependent)
constraints, such  degrees of freedom would be fixed by the
experimental bounds on the induced R-parity breaking terms.
As we have already used nine independent parameters to write the
lepton textures, the remaining two degrees of freedom can only affect
the  baryon number violating interactions,
 \be
 \lambda''_{ijk} U^c_i D^c_j D^c_k~.
\label{rb}
 \ee
In the above expression the effective coupling constant
$\lambda''_{ijk} \sim \e^{t_{ijk}}$, where the exponent $t_{ijk} =
\omega_i + \rho_j + \rho_k$. Now, we proceed to determine the
number of linearly independent parameters that give $t_{ijk}$.
First, we notice that $t_{ijk}$ contains only nine non zero
entries due to the discrete symmetry under $j\leftrightarrow k$
exchange, and to the color symmetry of quarks. There are also some
extra constraints that relate the different $t_{ijk}$  among
themselves, $t_{ikl}+t_{jkm}=t_{jkj}+t_{ikm}$; and with the
exponents in $Y_u$ and $Y_d$ textures through the symmetry
relationships: $t_{ikl}-t_{jkl}= n_{im} - n_{jm}$ and
$t_{ijk}-t_{ijl}= k_{ik} - k_{il}$, all for arbitrary family
indexes. Putting all these constraints together, it is easy to
realize that only one extra independent parameter is needed to
completely parametrize  the $\lambda''$ couplings. We then get \be
\lambda'' \equiv \left(\ba{ccc}
         \lambda''_{123} & \lambda''_{131} &\lambda''_{112}  \\
         \lambda''_{223} & \lambda''_{231} &\lambda''_{212}   \\
         \lambda''_{323} & \lambda''_{331} &\lambda''_{312}
       \ea\right)
      \sim \left(\ba{ccc}
          \e^{n_5-k_3}  & \e^{n_5-k_2} & \e^{n_5} \\
          \e^{n_4-k_3}  & \e^{n_4-k_2} & \e^{n_4} \\
          \e^{-k_3}     & \e^{-k_2}    &     1
       \ea\right)~\e^t ~.
\label{lrb}
\ee

Lepton and R-parity violating interactions appear in the MSSM due to the field
equivalence on lepton doublets, $L_i$, and the chiral Higgs superfield, $H_d$.
In fact, in the MSSM they carry the same quantum numbers
associated to the SM gauge group.
Thus, SUSY makes no distinction among them, and one
can write in the superpotential the interactions
 \be
 m_i L_i H_u + \lambda_{ijk} L_i E^c_j L_k + \lambda'_{ijk} L_i Q_j D^c_k~.
 \label{rl}
 \ee
However, those fields are actually different under the $U(1)_A$ symmetry.
At the level of the low energy effective theory, these terms will also be
generated by non renormalizable operators.
Moreover, the suppression on the different couplings appearing in
Eq.~(\ref{rl}) is governed by the lepton and quark textures. A simple
calculation allows to show that the bilinear mass coupling goes as
 \be
  m^i \sim Y_\nu^{i1} \e^{-r/2} M_P~.
 \ee
The fractional exponent $r/2$ means that such terms are only
accepted when $r$ is an even number.
For any other choice of this parameter this bilinear couplings will be
forbidden.
Notice, however, that  when r is a large negative number this
coupling is also zero (since then $Y_\nu =0$).
A similar conclusion comes out in the case of the other
two couplings, which are give as
 \be
 \lambda_{ijk}\sim Y_\nu^{i1}~ Y_e^{kj} \e^{-a-r/2}~;
 \qquad \mbox{ and } \qquad
 \lambda'_{ijk}\sim Y_\nu^{i1}~ Y_d^{jk} \e^{-a-r/2}~.
\label{lambda}
 \ee
This shows an interesting connection between the nature of
neutrino and R-parity violating interactions in the theory.
Actually, it reflects the well known conservation of lepton number
that appears when neutrinos are Dirac particles. Interestingly
enough, our mechanism is able  to generate  very light Dirac
neutrino masses, and at the same time to protect them from all
those potentially dangerous operators that could induce Majorana
masses.  Some clarification is needed at this point. The fact that
relates R-parity violating terms in Eq.~(\ref{rl}) with the
Majorana mass parameter $r$, that otherwise would not be expected
since $N$ does not appear at all on such couplings, is twofold.
First, the couplings in Eq.~(\ref{rl}) play no role in fixing the
$U(1)_A$ charges which at this point have been already defined by
all other couplings. Second, our assumption made on the  numbers
involved in the parametrization of fermion masses ($k,n,p$ and
$q's$), which we take as integers, makes $r$  the only possible
non integer piece in the involved combination of charges.

\section{A $SU(5)$ GUT embedding}

There are few remarks of interest
that one has to keep in mind while considering
the mechanism that we have analyzed in the previous sections.
Even though we have started with a theory that has twenty degrees of
freedom, represented by the assignment of $U(1)_A$ charges in
Eq.~(\ref{charges}), only nineteen independent parameters
($n_{1,...,5}$; $k_{1,2,3}$; $p_{1,...,5}$;  $q_{1,2,3}$; $r$; $a$ and $t$)
are actually needed to parametrize the effective low  energy theory.
Therefore, our present approach seems more economical.
It allows to describe the
physics of the low energy theory without bothering about
the actual charge distribution. Moreover, as we have one parameter less, there
is in principle a whole class of theories that may give same low energy physics.
We will use these extra degree
of freedom as a motivation to explore the
embedding of  our mechanism in the context of a $SU(5)$
unification  theory.

Let us consider a supersymmetric $SU(5)$ model where the MSSM matter content
is accommodated in the $\bar 5$ and $10$ representations as usual.
 \be
 \bar 5_i : (D^c, L)_i~;\qquad 10_i :(U^c, Q, E^c)_i~;\qquad
 \bar 5_d : H_d \qquad \mbox{ and }  5_u : H_u~.
 \ee
The
right handed neutrinos, as well as the $S$ superfield  are not originally in
the theory, thus we will add them  in the singlet representations:
$1_{N_i}$ and $1_s$, respectively.
As a comment, let us mention that the high energy fields needed to break the
$SU(5)$ symmetry, and to solve the doublet triplet splitting may be chargeless
under $U(1)_A$, such that their self couplings would not be affected. So that
pieces of the theory would remain unchanged.

Since $\bar 5$ and $10$ contain both  leptons and quarks, the
$U(1)_A$ charges [Eq.~(\ref{charges})] will be constrained. This
substantially reduces the number of degrees of freedom in the
theory to eleven, same that we will use to rewrite our
parametrization of mass matrices. First we write the Dirac mass
terms, $\bar 5~5_u 1_N$, and notice that no extra condition is
imposed on the parameters from these couplings, so $Y_\nu$ is
given as in Eq.~(\ref{yl}). On the other hand, as the up quark
masses now come from the couplings $10~10~5_u$,  the corresponding
mass matrix would be symmetric, and thus, its
parametrization has only three degrees of freedom. Hence, one has
to take $n_4=n_2$, and $n_5=n_3$ in Eq.~(\ref{yu}). In contrast,
now both charge lepton and down quark textures come from the same
generic coupling, $10~\bar 5~\bar 5_d$, which implies that $Y_d =
Y_e^T$.
Moreover, now, the most general form of the $Y_e$ texture
will only have a single degree of freedom, that enters as an
overall scale factor. Actually, one now should use $q_2 = n_2$ and
$q_3 = n_3$ in Eq.~(\ref{yl}). Finally, the right handed mass
terms, $M_P~1_N~1_N$, will remain as they were, and so our
conclusions regarding all other Majorana mass terms.

Therefore, in the $SU(5)$ context one gets the general parametrization
\bea
 Y_\nu \sim \left(\ba{ccc}
         1         & \e^{p_2 }    &\e^{p_3}\\
         \e^{p_4}  & \e^{p_2+p_4} &\e^{p_3+ p_4}\\
         \e^{p_5 } & \e^{p_2+p_5} &\e^{p_3+ p_5}
       \ea\right) \e^{p_1}~; &\qquad&
 \qquad ~
 Y_R \sim \left(\ba{ccc}
         1 & \e^{p_2 } &\e^{p_3}\\
         \e^{p_2} & \e^{2 p_2} &\e^{p_2+ p_3}\\
         \e^{p_3} & \e^{p_2+p_3} & \e^{2 p_3}
       \ea\right) \e^{r}~;\nonumber \\[1em]
Y_u \sim \left(\ba{ccc}
         \e^{2 n_3}     & \e^{n_2 + n_3} &\e^{n_3}\\
         \e^{n_2 + n_3} & \e^{2 n_2}     &\e^{n_2}\\
         \e^{n_3 }      & \e^{n_2}       & 1
       \ea\right) \e^{n_1}~; &\qquad&
Y_d^T; Y_e \sim \left(\ba{ccc}
         \e^{n_3 - p_5}     & \e^{n_2 - p_5}     &\e^{-p_5}\\
         \e^{n_3 + p_4-p_5} & \e^{n_2 + p_4-p_5} &\e^{p_4-p_5}\\
         \e^{n_3 }          & \e^{n_2}           & 1
       \ea\right) \e^{q}~.
\label{ysu5} \eea Ten independent parameters have been involved in
the above expressions: $n_{1,2,3}$, $p_{1,...,5}$, $q$ and $r$. As
before, the $\mu$-term, $\bar 5_d 5_u$, will fix one more, $a$.
Therefore, once the above parametrization are given, the theory
has no additional degrees of freedom. Any additional coupling on
the theory will be a prediction that can be written in terms of
the fermion mass textures. Such is the case of the R-parity
breaking terms,
 \be
 \lambda_{ijk}~\bar 5_i~10_j~\bar 5_k~;
 \ee
where the coupling constants are given as in Eq.~(\ref{lambda}):
$\lambda_{ijk}\sim Y_\nu^{i1}~ Y_d^{jk} \e^{-a-r/2}$.
It is worth noticing that the
baryon number violating interactions in Eq.~(\ref{lrb})
will now follow the same fate of all other R-parity breaking terms.
Interestingly, they all would
be forbidden in case neutrinos were Dirac particles,
which happens if one takes $r$ to be a fractional number.
Thus, in the $SU(5)$ context, the link between the nature of the
neutrino and the R-parity breaking terms becomes stronger.

 Lets make a final remark regarding the observation that
$Y_e=Y_d$ from above analysis. Last might be considered as a
negative observation. This is, however, a well known problem of
minimal $SU(5)$ GUT, for which solutions are also known. In order
to break the degeneracy among down quarks and charged leptons on
the second and third family, one can extend the theory by adding
an extra field, $\Sigma$, in the $24$ representation, which
carries no $U(1)_A$ charge and develops a VEV along the direction
$\langle\Sigma\rangle = V Diag(2,2,2,-3,-3)$. This field has the
coupling $10~\bar 5~\bar 5_{d} \Sigma$ that corrects the $Y_d$ and
$Y_e^T$ in a different way due to the different sign on the VEV.
Some small fine tuning in this case has to be introduced to get
the proper order of correction. Nevertheless, this will not affect
our other conclusions since the main hierarchy on the Yukawa
couplings would be given by similar expressions as before.

\section{Some Models with Dirac neutrinos and R-parity}

There is, in fact, a large number of possible tree level mass
matrices that one can generate from an additional $U(1)_A$
symmetry. Lets us consider some specific examples for illustration
proposes. The case of our interest would be that where the theory
has Dirac neutrinos, so, we fix  $r={1\over 2}$. To simplify, we
set $n_1=0$; $n_2=n_4=1$ and $n_3=n_5=2$ for $Y_u$ in Eq.~(\ref{ysu5}).
Also we take
$p_1= p+\ell$ and $p_{2,...,5}=-\ell$ for $Y_\nu$, with $\ell>p$.
That gives, up to order one factors, the quite appealing tree
level mass matrices
 \be
 M_u\sim m_t ~ \left(\ba{ccc}
         \e^4  & \e^3 & \e^2\\
         \e^3  & \e^2 & \e \\
         \e^2  & \e & 1
       \ea\right)~; \quad
M_e^T ; M_d
 \sim m ~ \left(\ba{ccc}
         \e^{\ell+2}  & \e^2 & \e^2\\
         \e^{\ell+1}  & \e & \e \\
         \e^\ell  & 1 & 1
       \ea\right)~; \quad
M_\nu \sim m_0 ~ \left(\ba{ccc}
         \e^\ell & 1 & 1\\
           1  & 0 & 0 \\
           1  & 0 & 0
       \ea\right)~.
\label{text} \ee  In  above expressions $M_e$ and $M_d$ are given
up to corrections that need to be added to break the $m_\mu/m_s=1$
condition, as we have already mentioned. Notice that we have also
reproduced the neutrino texture associated to models with the well
known $L_e - L_\mu-L_\tau$ symmetry~\cite{barbieri,bimax}. As we
expect neutrinos to be very light, in fact here $m_0=\e^pv$ has to
be $\sim\sqrt{\Delta m^2_{atm}}$, we are forced to take $p\sim
12$. That makes $\ell>12$. Thus, if no extra corrections are
considered, one gets for solar splitting $\Delta m^2_{\odot}\sim 4
\e^\ell\sim 10^{-9}~eV^2$, for $\ell=13$, which may be good for
oscillations in vacuum. Alternative explanations can be obtained
if the model admits  a soft hierarchy in the $h$ couplings, since
small departures from the texture would be introduced this way.

Another example. Let us instead take $p_1 = p + 2 \ell $ and the
same former values of other parameters. With this choice we get
the same mass matrices as in Eq.~(\ref{text}) but with the Dirac
neutrino texture
 \be
 M_\nu \sim m_0 ~ \left(\ba{ccc}
         \e^{2\ell} & \e^\ell & \e^\ell\\
           \e^\ell & 1 & 1 \\
           \e^\ell  & 1 & 1
       \ea\right)~.
\ee This generates a hierarchical pattern~\cite{barbieri}. Here
$\ell$ can take any value.  Many other textures are possible.

If the setup is the extended MSSM, only baryon number violating
interactions will appear. In the context of $SU(5)$ theory,
R-parity will be conserved. Assuming universality of soft SUSY
breaking terms at GUT scale, in the case SU(5) SUSY GUT,  or at
$M_{P}$ scale in MSSM case and supposing 
$m_{1/2}\geq 3 m_{0}$~\cite{kuzmin}, 
where $m_{1/2}$ and $m_{0}$ are common
gaugino and squark--slepton mass at GUT or at $M_{P}$ scale. Then,
the possible candidate for a LSP could  be the right handed
sneutrino, which can also play  the role of a dark matter
candidate. The phenomenology of this scenario has been studied  for
instance in Ref.~\cite{d2,kuzmin}.

\section{Conclusions and Remarks}

We have studied the generation of masses and mixings in models
with an anomalous $U(1)_A$ symmetry, which is supposed to be
spontaneously broken close to the string scale. In such models the
low energy interactions that are responsible for the generation of
masses and mixings and R-parity violation, in the context of an
effective MSSM, appear only at the renormalizable level due to the
non trivial assignments of $U(1)_A$ charges. By extending the MSSM
matter content via adding  three generations of right handed
neutrinos, which are singlets under the SM interactions, we have
uncovered  an intriguing link between R-parity breaking
interactions and Dirac neutrinos.

We have shown that in the context of the so extended MSSM, or
$SU(5)$ unification theory, the fermion mass textures can be
easily parameterized in terms of a small set of numerical
parameters. The parametrization has been  done in a way that does
not need the specification of $U(1)_A$ charge distribution among
low energy fields. The couplings are actually given in terms of
powers of a single parameter $\epsilon$. Thus, our
phenomenological parameters are represented by a set of exponents
of $\epsilon$ that appear on the effective couplings at low
energy. This phenomenological approach has the advantage not only
of predicting the most general form of the effective Yukawa
interactions, but also of relating the R-parity breaking couplings
with the Majorana mass textures and the hierarchy between weak and
Planck scales. From here, we showed that models with Dirac
neutrinos may only have  baryon number violating interactions,
which are governed by an overall scale. In $SU(5)$ extension of
the theory, R-parity appears as an exact symmetry provided that
neutrinos are Dirac particles. Some explicit examples of the tree
level textures produced for this class of models have been also
given. The mechanism may as well generate Majorana or pseudo-Dirac
neutrinos.

We would like also to mention  that our parameterizations are
equally good for theories where the textures are protected by a
global  $U(1)$ symmetry, which may be softly broken at some
intermediate scale $\Lambda$ by a scalar field $S$. In such a
case, it is obvious that in all our analysis one should take the
suppression in the non renormalizable operators to be given by the
scale of decoupling of the $U(1)$ theory, $\Lambda$, instead of
$M_P$. Hence, the expansion parameter $\e$ would be replaced by
the ratio
\[
\e\rightarrow {\langle S\rangle\over \Lambda}~.
\]
The only apparent changes will be suffered by
the overall scale of $M_R$ and the $\mu$-term that  explicitly
depend on the choice of $\Lambda$. No other important changes would arise.

Finally, we should mention that all the analysis in the present
paper has been done in the tree level approximation where one
takes all dimensionless couplings, $h_{u,d,e,\nu}$, to be one, according to 
the philosophy of effective field theories.
Small deviations on our effective Yukawa couplings could be
expected, though they will not affect 
the textures provided that not large hierarchies are introduced
through $h_{u,d,e,\nu}$.


\acknowledgments

We gratefully acknowledge the helpful discussions with: K.S. Babu,
Borut Bajc, Stefano Bertolini,  Ara Ioannisian,  Goran
Senjanovi\'c, Marco Serone and  Alexei Smirnov. We also would like
to  thank  A.S. Joshipura, R. Vaidya and  S. Vempati for comments.
This work was partially supported by EEC under the TMR contracts
ERBFMRX-CT960090 and HPRN-CT-2000-00152. The work of I.G. has been
partially supported in part by DOE Grant \# DE-FG03-98ER-41076.

\end{document}